**Directing Avatars in Live Performances - an Autonomy Simulacrum of Virtual Entities**

Georges Gagneré, INREV-AIAC, Paris 8 University

**Abstract**
We propose to review the main stages in the computer history of virtual actors, with a view to the exploration of virtual reality and discussion on different approaches to human simulation. The notion of autonomy emerges as a key issue for the virtual entities. We then explore one way of building elements of autonomy and conclude with an example of avatar stage direction leading to a simulacrum of autonomy in a live performance.

## Introduction

Artificial beings have fascinated natural humans ever since they discovered they could transform nature using tools. In his latest book *Automates, robots et humains virtuels dans les arts vivants (Automata, robots, and virtual humans in the performing arts)*, Edmond Couchot recalls that Homer in the *Illiad* evoked "mechanical maids, made of gold, having the appearance of young girls and created by [Hephaestus] to be helped in his work, all endowed with speech, reflective thought (*noos*) and a certain physical strength" (Couchot 2022, 16)[1]. The divine blacksmith had thus invented creatures intelligent enough to assist him. Three millennia later, the project is being carried out by humans themselves, thanks to the recently invented science of computing and the sub-discipline of artificial intelligence, a complex field that aims to completely externalize our cognitive processes.

In the first part of this chapter, we propose to review the main stages in the history of computer-based virtual actors, with a view to the emergence and exploration of virtual reality. This will enable us to discuss the various approaches that have been taken to human simulation, i.e. the dynamic reproduction of complex human behaviours and appearances in the form of models, algorithms, and programs. The notion of autonomy emerges as a key issue in the creation of virtual entities. The second part is then devoted to exploring one way of building elements of autonomy by explaining how the virtual actors used in video games and performing arts can be moved. We break down the various techniques used to control them and simulate elements of autonomy in virtual worlds. Finally, we focus on an example of avatar stage direction leading to a simulacrum of autonomy in a live performance.

## The simulation of the virtual actor

The first realistic virtual humans were the result of an artistic commission to a couple of computer science researchers, Nadia and Daniel Thalmann, originally from Switzerland, who began their career in Canada in 1977. After developing the MIRA-3D computer language for computer-generated images, with which the short film *Dream Flight*, directed by Philippe Bergeron, won international scientific acclaim in 1982 (Thalmann, Magnenat-Thalmann and

---

[1] Unless otherwise indicated, this and all other citations from non-English sources have been translated by the author.

Bergeron 1982), they made the film *Rendez-vous à Montréal* (Magnenat-Thalmann and Thalmann 1987). This film, whose subject matter is a meeting between a virtual Marylin Monroe and Humphrey Bogart at Montreal's Bon Secours market, was screened in 1987 for the 100th anniversary of the Canadian Order of Engineers. The realism of Virtual Marilyn, representing the culmination of ten years of research in computer graphics, made a lasting impression. This type of work was the first concrete evidence of the possibilities that digital technology and computer language offer for the simulation of reality.

### *Inhabiting the new virtual reality*

The late 1980s saw the emergence of what Jaron Lannier called Virtual Reality in 1989 (Heim 1994, xvii), which offers natural humans sensory access to a 3D simulation of physical reality *via* a virtual reality headset and data gloves. Before the term was coined, Myron Krueger paved the way by using the term Artificial Reality to refer to the artistic experiments he carried out in the 1970s (Krueger 1983), particularly the experiment *Videoplace* (1974), which allowed viewers to inhabit a computer-generated image with their digital silhouette and to interact with artificial creatures. It became possible to immerse oneself in a digital pixel matrix, manipulate it in real time and tame the entities inhabiting it.

Virtual reality turned on its head the bold proposals Brenda Laurel had made to the developer community in *Computers as Theatre*, to use Aristotle's reflections on Greek tragedy in his book *Poetics* as a model for building efficient interfaces (Laurel 1993, 50). Drawing inspiration from the process of *mimesis*, and its dramaturgical mechanisms for describing actions in the form of tragedy triggering *catharsis*, she proposed the development of effective, stimulating interactions of a user with the actions proposed by a program. However, this controlled, vertical approach to an interaction scenario soon appeared limiting to Laurel, not least because of the immeasurable, unpredictable richness offered by emergent virtual reality. The key notions of interactivity and simulation allow users to explore new realities by inhabiting the envelopes of artificial creatures. Inspired by *Videoplace*, Brenda Laurel conceived and realized *Placeholder* in 1992, one of the first performative art installations in virtual reality, in which two spectator-performers project themselves into the image of a raven, a spider or a snake to experience new bodily sensations.

### *Simulating the natural human*

Nadia and Daniel Thalmann returned to Switzerland after their successful experiments in Canada and founded respectively the MIRALab at the University of Geneva and the Computer Graphics Laboratory at the Ecole Polytechnique Fédérale de Lausanne (EPFL) in the late 1980s, where they focused on 3D simulation of the human being. Two possible approaches soon became clear: on the one hand, enabling users to project themselves into cyberspace with an appearance that resembled them, connecting with other virtual human beings; and, on the other, populating the virtual world with artificial humans who would possess their own autonomy, thanks to research into artificial intelligence. In the late 1990s, their work led to the first classification of virtual humans[2]. In this schema, the avatar serves as a simple artificial vehicle for a human user, enabling him or her to immerse and exist in a manner close to his or her natural bodily vehicle in the new reality simulated by computers and their peripherals. Jean-Claude Heudin also notes a reversal of polarities in relation to the etymology of the notion, which comes from the Sanskrit *Avatāra*, and which in the Hindu religious tradition designates the incarnation of an extra-human entity in a living earthly body. But in the context of virtual reality, "instead of a deity incarnating into a material body, it's a material body that 'disincarnates' into a virtual representation." (Heudin 2008, 210).

The Thalmanns' project enables a participant to inhabit his or her own synthesized body envelope with the aid of a motion capture system, and to visualize the result directly in the virtual environment with a 3D head-mounted display. The goal is not necessarily for the participant to see himself, but rather to appear as his expressive digital double to another participant. Conceivably, this makes it more convenient for the participant to continue to exist in a 3D environment, without the need to wear a motion capture suit and headset. "Disembodiment" in the avatar is then achieved by means of control peripherals akin to the bars used to move the wire of a marionette, although guidance is generally very different from that of a puppet, as we shall see later. In a further logical step, guidance can be automated by algorithms, giving the virtual characters elements of autonomy. As Daniel Thalmann puts it: "Our ultimate goal is to create credible, realistic synthetic actors - intelligent, autonomous virtual humans with adaptability, perception and memory. These actors should be able to act freely and emotionally. Ideally, they should be conscious and unpredictable" (Thalmann 1998, 65). To achieve this, the project needs to draw on the computer science discipline of artificial intelligence, which has seen regular ups and downs in institutional funding since the 1960s. The avatar's virtual body is no longer inhabited; it must exist on its own, and to do this, it must be equipped with mechanisms that enable it to move, perceive the virtual world and other virtual entities, feel emotions and enter into social relations. At the 1997 Interactive Telecom conference, this research project gave rise to a tennis match in which a player in Geneva at MIRALab played with his avatar against the avatar of another player at EPFL in Lausanne, while the Virtual Marilyn from the 1987 project refereed the match autonomously.

### *Different approaches to simulation*

There is as yet no standard for simulating virtual humans, and every laboratory, video game engine or cyberspace platform, every researcher and every artist has their own simulation model or approach. Simulation based on the natural laws of physics and biology led Jeffrey Ventrella to propose a model (Ventrella 2000) that could evolve according to the laws of genetics, and which was subsequently used to develop the online virtual world *Second Life* from 2003 (Linden Lab 2003). Ken Perlin's work led him to combine technology with an artistic approach to create virtual characters capable of improvisation (Perlin 2000). This enabled him to generate real-time theatrical performances in which the characters could interact with physical participants, lasting up to a week without interruption.

---

[2] This classification contains four categories, avatars, guided actors, autonomous actors, and interactive-perceptive actors (Cavazza et al., 1998), that we detail in the section *Describing the autonomy of a virtual actor*.

Ventrella and Perlin's work was inspired by Norman Badler's 1980s research on the human simulation project *Jack*, which was used, in particular, to prototype the NASA's Space Shuttle. *Jack* generalized the principle of inverted kinematics, enabling the joints of a virtual human to be manipulated in a way that is plausible for contact with the virtual environment (Badler, Barsky, and Zeltzer 1991). It was under Badler's direction that Catherine Pelachaud submitted her PhD thesis (Pelachaud 1992), before pursuing her research into socio-emotional conversational agents. This scientific discipline works on giving a human appearance to entities endowed with artificial intelligence, enabling them to interact with natural humans in a wide range of social activities. Catherine Pelachaud has worked, in particular, on the simulation of emotions and non-verbal expressions and created the GRETA virtual agent model (Grimaldi and Pelachaud 2021).

In parallel to these approaches, which were all based on the implementation of a pre-existing model endowed with behavioural skills that enable it to simulate social interactions, another approach inspired by what the history of cognitive sciences describes as the fourth wave of cybernetics, starting with the work of Humberto Maturana and Francisco Varela (Maturana and Varela 1987), has led researchers to develop autonomy in virtual humans based on the concept of emergence. In the early 2000s, mathematician and artist Michel Bret built virtual dancers who learn to dance on their own using neural networks (Bret 2000). This has enabled him to create choreographic performances with the artistic collaboration of Marie-Hélène Tramus at the Digital Image and Virtual Reality (Image Numérique et Réalité Virtuelle) laboratory at Paris 8 University (Bret, Tramus and Berthoz 2005). These experiments involve what they call "second interactivity" (Couchot, Tramus and Bret 2003). They depart from the action-reaction processes typical of early cybernetic theories to rather approach presence and interaction in line with the concept of enaction developed by Varela, which is based on "the enactment of a world and a mind on the basis of a history of the variety of actions that a being in the world performs." (Varela, Thompson and Rosch 1981, 9)

This approach to autonomy in artificial intelligence was formalized by Jacques Tisseau at the Centre Européen de Réalité Virtuelle (CERV) in Brest. It combines the reciprocal influence of the notions of prediction, action and adaptation (Tisseau, Parenthoen and Harrouet 2006). This led CERV to develop the Atelier de Réalité Virtuelle (ARéVi) programming library, which was later used by researcher Pierre de Loor to create a experimental dialogue between an autonomous entity and an actress. Conducted as part of the ANR INGREDIBLE (Bevacqua et al. 2016) project in 2010, in collaboration with a theatre company, the experiment led the researcher to conclude that "the user's engagement can be improved if the virtual actor has behavioural characteristics defined through enaction" (De Loor et al. 2014).

At the same time, the fifth and final volume of the *Traité de réalité virtuelle* launched in 2003 by Ecole des Mines researcher Philippe Fuchs is entirely devoted to virtual humans and offers a panorama of advances within several scientific disciplines that are brought together to collaborate on interdisciplinary research projects (Fuchs, Moreau and Donikian 2009). Production of this volume was coordinated by Stéphane Donikian, whose research into interactions in crowds is the basis of the Goalem tools for simulating complex autonomous behaviours used in many fields, including video games and animated films (Goalem 2023). This research is continuing and benefiting from continuous advances, with the 2020s marking the rise of approaches based on deep learning and new algorithms whose results in the recognition and simulation of language and images maintain the fascination of natural humans for the subject[3].

### Describing the autonomy of a virtual actor

We will now return to Daniel Thalmann's classificatory schema (see note 1) and its reinterpretation by artist-researcher Cédric Plessiet to question the relationship between the avatar and the autonomous virtual human. In the field of video games, digital art installations and the performing arts, Plessiet assumes that the movements of virtual entities can be analysed along two axes: that of the origin of movement, i.e. the principle that leads the entity to carry out gestures and actions in the virtual world with its virtual body; and that of the decision to move, i.e. the choice and intention to carry out a movement, and which corresponds to the possession, or not, of free will (Plessiet and al. 2019). He then deduces four types, depending on whether the origin and decision belong to the virtual entity or not (are internal or external), and which he names after theatrical or mythological notions. For the Virtual Puppet, the origin and decision of its movements lie outside itself. It corresponds to what Thalmann called the Avatar. By contrast, the Virtual Actor generates and decides on movements, corresponding to Thalmann's Interactive-Perceptive Actor. The entity that decides on its movements but does not generate them is a Virtual Mask, in reference to the theatrical tradition of masks, in which the performer must adapt his performance to the personality of the mask he wears and which "controls" him. Finally, the last type is the Virtual Golem, who is at the origin of his movements but a slave to external decisions[4]. Thalmann's Guided Actor and Autonomous Actor fall into this category in the sense that they must possess their own means of movement to respond to the instructions from an external human guide or from environment detection algorithms, which we can then consider, following Plessiet, as an internal stratum of movement animation.

Thalmann's classification was influenced by the discovery and exploration of virtual reality, which at the end of the 1990s was still confined to those research laboratories that possessed powerful graphics workstations and access peripherals that were still very expensive: motion capture systems, data gloves, HMD or CAVEs (rooms where 3D images are projected onto the walls and the user wears special glasses) (Cruz-Neira and al. 1993). The challenge of virtual reality is to enable the human being to access this new reality via an Avatar in an environment that can simulate social activities with the other three types of Actors. The ultimate goal of the research is to simulate all human complexity in the digital matrix to achieve the figure of the Interactive-Perceptive Actor. Plessiet's classification focuses on the virtual entity from the point of view of its computer construction and enables us to recognize and anticipate the constraints that appear depending on the situation in which the entity is immersed, in the field of cultural and artistic uses.

---

[3] The publication *Actor & Avatar* (Mersch and al.2023) gives references providing an overview on multiple research projects, results and topics related to virtual actor and the future of AI.

[4] The Golem refers to the mythological figure from Jewish folklore that is formed from earth, brought to life by a magical incantation and who blindly obeys its master, without free will.

In both cases, we observe a common strategy of describing a simulation or simulacrum of the Actor (Perceptive and Interactive) by making the Puppet (the Avatar) climb all the stages of a progressive emancipation on the way to the Golem (the Guided or Autonomous Actor). This is radically different from the enactive approach taken by Bret, Tisseau and De Loor. However, it is widely used in the fields of animation, video games and the performing arts, and will allow us to describe several steps in the construction of Golems aimed at approaching a simulacrum of the Virtual Actor.

# From the Puppet to the Actor *via* the Golem

## *Sculpture-matrix and real time*

Our starting point is Plessiet's notion of the sculpture-matrix (Plessiet 2019, 28) as a minimal computer representation of an anthropomorphic figure, which can be seen as an echo of the notion of the image-matrix proposed by Edmond Couchot in the late 1980s to describe the revolution that was about to overturn artistic and cultural practices (Couchot 1988, 189-219). The image-matrix implied abandoning the analogue concept of the optical image in favour of the expressive possibilities inherent in the nascent digital order, which made an image the association of a new technology of perception of the real/physical and a computer language with precisely established rules. The sculpture-matrix proposes an even more precise appropriation of all the potential offered by computer simulation.

Figure 1 (A and B) shows two sculpture-matrices of me: on the left (1A), a 3D representation of my body and face created by Plessiet. It is attached by the process of skinning to a virtual armature of joints and sticks, the union of the two being called bones and forming the skeleton of the sculpture-matrix. Describing the positions and rotations of the skeleton's bones animates the entity. The image on the right (1B) shows the same skeleton dressed by another skeletal mesh, which has the shape of an almost flat silhouette, except for the feet, of a previous version of my avatar. The field of possibilities is infinite, as is the range of positions the sculpture-matrix can take in 3D space.

Animation and video games use two main techniques to animate a sculpture-matrix. The oldest involves equipping the bones of the skeleton with controllers to form a rig and creating the movement curves of each bone from successive poses (see figure 1C). The artist-animator uses the rig to achieve the desired poses. The other technique is to use a motion capture system to calculate the positions and rotations of the physical limbs of the person performing the movement. This is a complex operation, as it involves reconstructing an intermediate skeleton from the capture data, and then matching this skeleton with the skeleton of the sculpture-matrix, while respecting the proportions, using a motion-retargeting process. Both techniques make it possible to save interpolation data between animation poses or successive motion sensor positions and replay them at will. Motion curves can then be edited and modified to adjust motion quality.

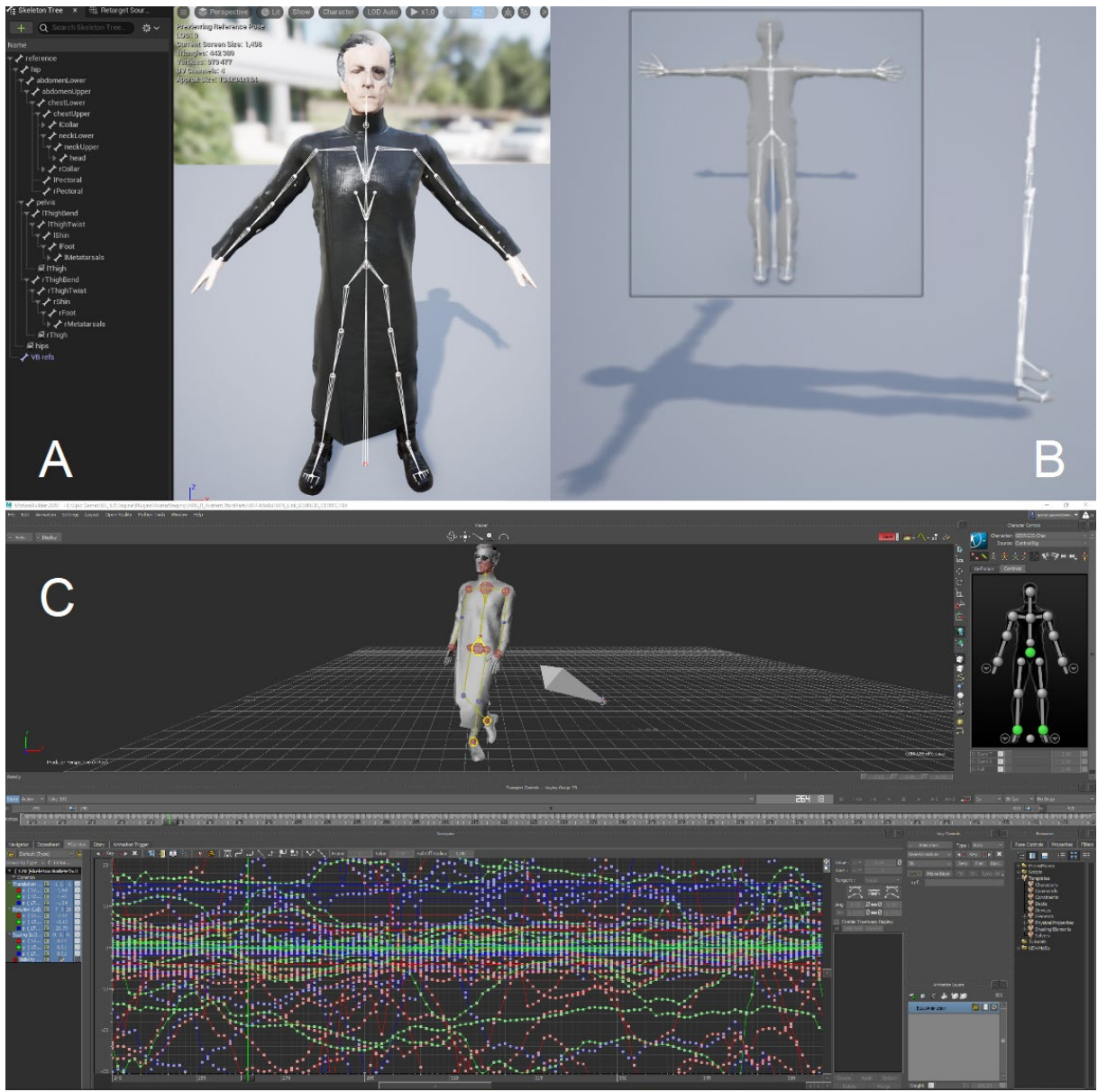

*Figure 1 : Two sculpture-matrices with their skeleton in Unreal Engine (A and B) (Epic Games 2023) and the motion curves in Motion Builder (Autodesk 2023) (C). Screenshot by Georges Gagneré*

At this stage, we make a distinction between the field of animation, which uses precalculated computer techniques, and that of video games, which uses real-time techniques to obtain the final images that are presented to the viewer/player. In precalculated mode, each image is calculated separately, with computing times depending on the complexity of the VFX and the rendering quality of the sculpture-matrix movements. The result is put in a file and transmitted linearly to the viewer during the broadcast. What we are interested in here is the real-time mode of using motion information to deploy interactions with the recipient of the images – wether a player, actor or active spectator. Our starting point is the control of an avatar by an actress wearing a motion capture suit, whom we will call a mocaptress. We are in the mixed-reality scenic environment proposed by the AvatarStaging setup (cf. figure 2) (Gagneré and Plessiet 2018).

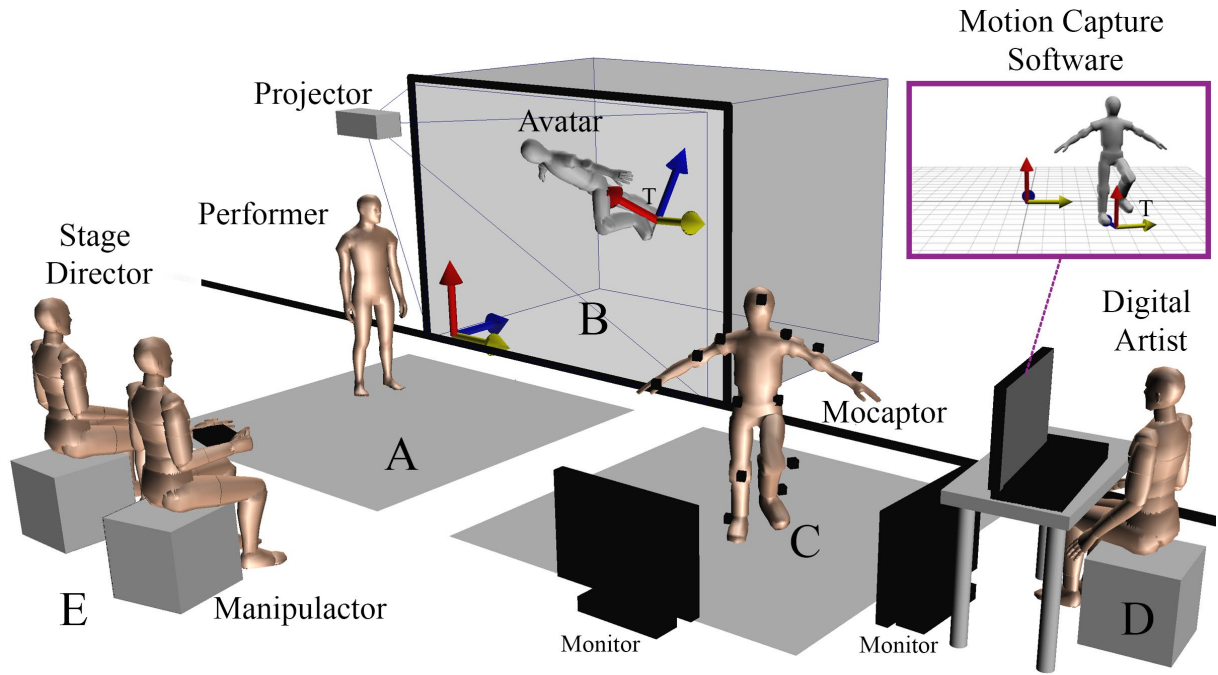

*Figure 2 : AvatarStaging setup showing a mocaptor controlling his avatar by motion capture. Image by Georges Gagneré*

The mocaptress moves in her own acting space C, surrounded by monitors that allow her to adapt the position of her avatar in space B to its physical partner in space A. She keeps eye contact with the avatar in every position so that to fine-tuning the staging situation. We propose to imagine that there are no monitors and that the mocaptress must remain oriented towards the video rendering screen of virtual space B to control its image. This scenario has already been encountered in artistic productions and introduces the concept of guiding an avatar (Gagneré and Plessiet 2019). Since we are in real time, we can also modify the avatar's position without the mocaptress moving.

### Game development tools

We can use a method that consists of placing the avatar in a capsule, transforming its movement to make it walk on the spot and simultaneously moving the capsule along the trajectory of a reference that corresponds to the root of the avatar (cf. figure 3A.). This root corresponds to the projection of the skeleton's main bone, in relation to which all other bones are referenced using the parentage method. This technique of using real-time motion from a motion capture system or from playing a pre-recorded animation is called root motion. Figure 3A shows two states of movement in which we can recognize the character in an idle position on the left, and the character walking in place in front of the white door. Combining these two movements in a black wireframe capsule allows the character to walk by orienting the capsule in space. With this technique, to achieve the avatar's previous orientation according to the actor's movement, all you must do is turn the capsule on the spot at the right angle.

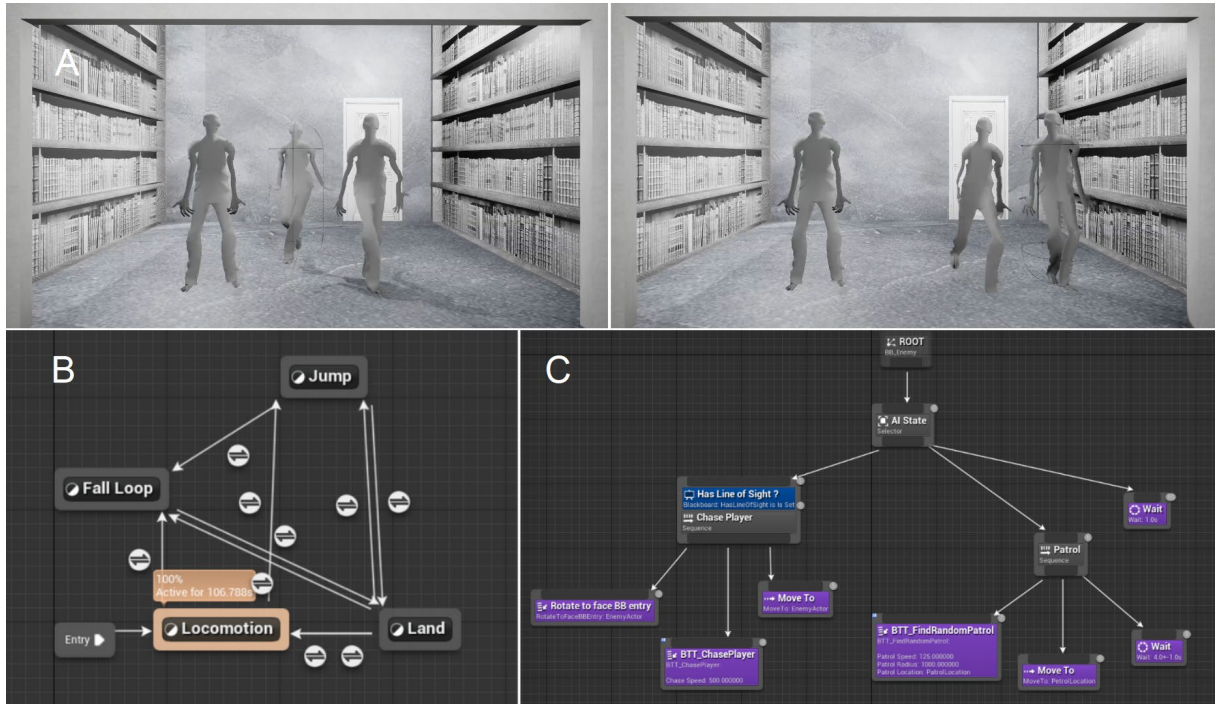

*Figure 3: Avatar root motion using a capsule (A). Finite State Machine (B). Behaviour Tree (C) (Epic Games 2023b). A: Image by Georges Gagneré; B and C: Images by Epic Games© public online tutorial.*

In a live-performance setting, when working with a mocaptress, these techniques are not necessary, as you can rely on her skill to perform the avatar's desired movements. By contrast, they are necessary in a video game context. Apart from the non-interactive moments of the cinematics, which are animations and mobilize pre-recorded files, it is not possible to use an animation that moves a player's avatar from a given point to another along a trajectory. Instead, you need to use of a Virtual Golem/Guided Actor. The player is free to move and orientate the capsule at any speed in 3D space, and the avatar walks in place, adjusting its speed. This technique gives correct visual results, as long as no attention is paid to the position of the feet, which glide lightly over the ground before stabilizing when they reach the right speed. The algorithm used to control the movements is a Finite State Machine (FSM), which allows different movements to be chained together according to precise rules that corresponds to the movement possibilities offered to the player (see figure 3B).

In what way can we consider an entity that uses a walking animation combined with an idle animation to be a Golem, since it has been transmitted animations produced by a mocaptor or animator, i.e. an outside person? Precisely because, unlike a Puppet, the entity is equipped with a mechanism that it activates with relative autonomy according to instructions transmitted to it by an external player. It could be argued that the FSM is also programmed by a developer outside the entity. We would then have to consider the Golem as a kind of pre-programmed automaton. The assessment is subjective and linked to the evaluation of the complexity of the algorithms associated with the entity. An example often used is the pathfinding mechanism, which consists of analysing the environment via a NavMesh (navigation mesh) and applying an algorithm to move from one point to another, avoiding any obstacles in 3D space. In this case, we consider that the start and end points are given by external instructions, but that the organization of the movement based on the pathfinding algorithm and the NavMesh enables the entity to use its walking capsule to generate a movement that will bypass the obstacles encountered without external guidance. The entity is thus the originator of its own movements.

The next step is to increase the Golem's degree of autonomy by equipping it with a Behaviour Tree (BT) mechanism, which consists of assembling FSMs that react to changes in parameters describing the 3D environment (see figure 3C). The Golem's behavioural skills progress in the direction of those imagined for the interactive-perceptive actor. FSM, pathfinding, NavMesh and BT are video game engine tools that can be used to endow an entity with elements of artificial intelligence. Will the increasing complexity of these tools lead to conscious autonomy? Proponents of the enactive approach would say no, and video game developers or artists are not seeking to achieve this goal. Rather, they aim to materialize simulacra of autonomy that elicit a "willing suspension of disbelief" (Coleridge 1817, 145) from players and viewers alike, so that they can immerse themselves in a game or interactive installation.

### *Autonomous actor simulacrum: The Shadow*

We conclude this exploration of virtual actors by describing a simulacrum of autonomy used in a work combining sculpture-matrix, video game tools, theatre and virtual reality on a mixed stage. *The Shadow* (cf. figure 4) is a theatrical performance I conceived and produced in 2019 involving a physical actor and five virtual entities that I call shadow avatars, in reference to their nature as flat silhouettes that can be inhabited by a mocaptor (cf. figure 1B). My approach to the avatar was initially based exclusively on the use of real-time motion capture using AKN_Regie software, which I developed as a plugin in the Unreal Engine video game engine environment (Gagneré 2023). This software combines the engine's resources to apply them to what I call avatar direction to achieve a simulacrum of autonomy. In broad terms, avatar direction applies theatrical actor direction methods to a mocaptor controlling an avatar in such a way that it performs replayable and modifiable scenic gestures according to the acting of its physical partner.

I imagined the presence of shadow avatars in relation to a physical actor who tells Andersen's fairy tale *The Shadow*, in which a scientist abandons his shadow after sending it to explore a house inhabited by a mysterious musician, who turns out to be Poetry. In contact with her, the scholar's shadow becomes almost human and grows rich, using its shadow powers to sneak into the homes of the wealthy and blackmail those with secrets to hide. After several years, the shadow returns to torment its former owner. I imagined that an actor would tell the tale to both spectators and shadow avatars on stage in a virtual small theatre. As the story unfolded, the shadow avatars, captivated by it, start miming the protagonists, reliving the various episodes over the course of the 50-minute show (Gagneré 2020).

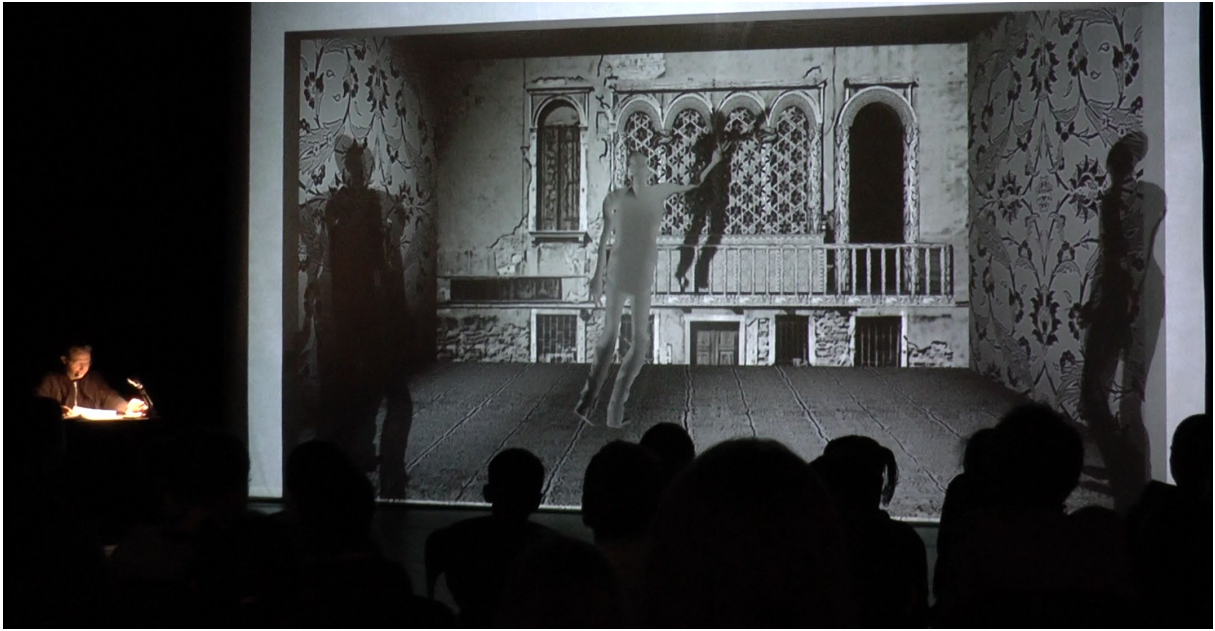

*Figure 4:* The Shadow *(2019), after Andersen, direction Georges Gagneré with Eric Jakobiak. Image by Georges Gagneré*

Each shadow avatar's journey consists of a series of cues triggered directly by the physical actor reading the story, seated at a small table to the left of the virtual theatre, using a midi controller (see figure 4). There is no stage manager standing between the actor and the avatars. The paths of the avatars form a kind of visual score of pre-recorded animations, with the performer as narrator. Each animation was carefully recorded by a mocaptor listening to the tale's narration under my stage direction, according to the following acting protocol: we systematically alternated stage gestures that I call salient, followed by non-salient gestures. This intuitive approach to stage salience is inspired by the principle of idle loops, used in video game animation. This salience is the opposite of an idle gesture, which can be looped back on itself without creating a conspicuous repetition effect. An example is the resting position prior to walking, represented in figure 1B by the character on the left. He assumes a waiting position on the spot, not very active but not immobile, curled in on himself while waiting to receive the order to walk, still on the spot but in the capsule set in motion. A salient gesture, on the other hand, is expressive and cannot be looped without the spectator immediately detecting the repetition. The threshold of salience is an empirical fact, and a very attentive observer could detect a looping process, even with a non-salient idle animation.

Based on this principle of systematic alternation between salient and non-salient stage gestures for all recorded animations, I developed the Salient-Idle Player feature in AKN_Regie, which equips each shadow avatars and allows the actor to successively trigger the acting animation, ending in a loop where the avatar listens and waits for further instructions. Does this feature transform the shadow avatar from puppet to "basic" golem? Probably not. But it's a puppet guided by an internal mechanism specific to the nature of its movements, and which receives its acting instructions from the external actor. The end result, from the spectator's point of view, is to trigger the suspension of disbelief that makes him feel that the five shadow avatars are performing the show autonomously in relation to the physical narrator.

## Perspectives

> The human lineage exists only insofar as it stakes its own existence on a kind of headlong rush with no horizon - other than the one it draws for itself - for some, or towards an end that transcends it for others. Hence the impossibility of defining it in an absolute manner. Hominization never ends. Artificial living arts take part in this debate and suggest, by tinkering, a possible way of living together with these newcomers. (Couchot 2022, 205)

From the first stylized simulations of *Dream Flight* to the tinkering with shadow avatars in *The Shadow*, it is clear that the field of virtual actor simulation offers an infinite territory of exploration for artists. For scientists, it is a question of simulating human beings and offering them new ways of inhabiting virtual reality. In the wake of Georges Simondon, who demonstrated that culture and the arts must appropriate technical objects as bearers of humanity, to actively accompany the evolution of our societies (Simondon 1958), I continue to explore new avenues of autonomy for shadow avatars, gradually integrating FSM and BT into AKN_Regie. I wish to build performative experiences that will enable spectators and mocaptors to meet these new creatures immersed in virtual reality. Instead of remaining a mere spectator of shadow avatars, I wish to experiment with the possibility of one's own shadow escaping out of control and improvising with other shadows that have also managed to elude their owners. This could lead the dispossessed owners to make strange discoveries. To extend Couchot's vision of virtual newcomers living with natural humans, it would then be a matter of exploring new potentialities of presence and relationship to others.

Couchot, Edmond, Marie-Hélène Tramus, Michel Bret, and Diana Domingues, eds. 2003. "A segunda interativida. Em direção a novas práticas artísticas". In *Arte e vida no século XXI. Tecnologia, ciência criatividade*. São Paulo: Unesp.

Cruz-Neira Carolina, Daniel J. Sandin and Thomas A. DeFanti. 1993. "Surround-Screen Projection-based Virtual Reality: The Design and Implementation of the CAVE", *SIGGRAPH'93: Proceedings of the 20th Annual Conference on Computer Graphics and Interactive Techniques*, 135–142.

De Loor, Pierre, Kristen Manac'H, Charlie Windelschmidt and Frédéric Devillers. 2014. "Connecting Interactive Arts and Virtual Reality with Enaction." *Journal of Virtual Reality and Broadcasting*, 11, no. 2 (February). https://doi.org/10.20385/1860-2037/11.2014.2.

Epic Games. 2023. "Unreal Engine" Accessed May 28, 2023. https://www.unrealengine.com.

———. 2023b. "Unreal Engine Documentation" Accessed May 28, 2023. https://docs.unrealengine.com.

Fuchs, Philippe, Guillaume Moreau, and Stéphane Donikian. 2009. *Le traité de la réalité virtuelle. Volume 5: les humains virtuels*. Paris: Presses des MINES.

Gagneré, Georges. 2020. "*The Shadow.*" *Proceedings of the 7th International Conference on Movement and Computing (MOCO '20)*, Association for Computing Machinery, New York, NY, USA, 2020, Article 31, 1–2. https://doi.org/10.1145/3401956.3404250.

———. 2023. "AKN_Regie, un plugin dans Unreal Engine pour la direction d'avatar sur une scène mixte. " *Journées d'Informatique Théâtrale – JIT22*, October, Lyon.

Gagneré, Georges, and Cédric Plessiet. 2018. "Experiencing avatar direction in low cost theatrical mixed reality setup." *Proceedings of the 5th International Conference on Movement and Computing (MOCO '18)*, Association for Computing Machinery, New York, NY, USA, Article 55, 1–6. https://doi.org/10.1145/3212721.3212892.

———. 2019. "Espace virtuel interconnecté et Théâtre (2). Influences sur le jeu scénique." *Revue: Internet des objets* 3, no. 1 (February). https://doi.org/10.21494/iste.op.2019.0322.

Gagneré, Georges, Tom Mays, and Anastasiia Ternova. 2020. "How a Hyper-actor directs Avatars in Virtual Shadow Theater." *Proceedings of the 7th International Conference on Movement and Computing (MOCO '20)*, Association for Computing Machinery, New York, NY, USA, 2020, Article 15, 1–9. https://doi.org/10.1145/3401956.3404234.

Goalem. 2023. "Goalem tools" Accessed May 28, 2023. https://golaem.com/.

Grimaldi, Michele, and Catherine Pelachaud C. 2021. "Generation of Multimodal Behaviors in the Greta platform." *21st ACM International Conference on Intelligent Virtual Agents*, Sep 2021, Kyoto (virtual), Japan. htpps://doi.org/10.1145/3472306.3478368.

Heim, Michael. 1994. *The Metaphysics of virtual reality*. New-York: Oxford University Press.

Heudin, Jean-Claude. 2008. Les créatures artificielles. Des automates aux mondes virtuels. Paris: Odile Jacob.

Krueger, Myron W. 1983. *Artificial Reality*. Boston: Addison-Wesley.

Laurel, Brenda. 1993. *Computers as Theater*. Boston: Addison-Wesley.

Linden Lab. 2023. "Second Life" Accessed May 28, 2023. https://secondlife.com/.